\documentclass[showpacs,apl,twocolumn,floatfix]{revtex4}
\usepackage{graphicx,float,amsmath}
\usepackage{mathrsfs}
\usepackage{natbib}
\usepackage{amsmath}

\begin{document}
\date{\today}

\title{Effect of a uniaxial strain on optical spin orientation in  cubic semiconductors}

\author{T. Amand$^1$}
\author{D. Paget$^2$}

\affiliation{%
 $^1$ Universit\'e de Toulouse, INSA-CNRS-UPS, LPCNO, 135 Av. Rangueil, 31077 Toulouse, France}

\affiliation{% 
$^2$ Physique de la mati\`ere condens\'ee, Ecole Polytechnique, CNRS, Universit\'e Paris Saclay, 91128 Palaiseau, France}

\begin{abstract}
The effect of a uniaxial strain on the optical spin orientation of a  cubic semiconductor  is investigated by calculating the valence wavefunctions, the optical oscillator strengths and the initial electron spin polarization for near resonant light excitation from heavy and light valence levels.  A strain orientation along the  $[0 0 1]$, $[1 1 1]$ or $[\bar{1} 1 0]$  crystal direction  and a circularly-polarized light excitation parallel or perpendicular to the strain are considered. For all these cases, the total conduction electtron spin polarization has a universal character since i)   the oscillator strengths do not depend  on the magnitude of the strain but only on its sign.. ii)  Although the oscillator strengths strongly depend on the configuration, the conduction electron spin polarization generated by optical transitions from both  the heavy and light valence levels induced by $\sigma^+$ light excitation  is in all  cases equal to $-0.5$ as predicted by the simple atomic model. iii) The spin polarization generated by light excitation from heavy and light valence levels does not  depend on  the  deformation tensor  and on the valence deformation potential, except  for  a strain along $[\bar{1} 1 0]$. \
\end{abstract}
\maketitle

\section{Introduction}
 Since the discovery of the optical spin orientation of conduction electrons in semiconductors   \cite{zakh1984},   a large number of works have investigated the spin properties of carriers. The spin orientation has been shown  to be strongly affected by a uniaxial strain, which splits the heavy and light valence levels and   increases the hole spin relaxation time \cite{roussignol1992, linpeng2021, egorov2005, tessler1994, marie1999}. \ 
 
However, in spite of a reflexion over several decades, the factors which influence the spin orientation of conduction electrons at the instant of creation are still not completely clear.  The vast majority of the  studies  consider transitions between atomic levels. This model implies  that optical spin orientation occurs by transfer of momentum from the photon to the valence hole, thus leading, for a $\sigma^+$-polarized light excitation,  to an initial polarization of $P_i^h=-1$ for electrons generated from the heavy valence level , a polarization  $P_i^l=+1$ for electrons generated from the light  valence level and to an  overall  initial spin polarization at creation of  $P_i=-0.5$. This model has been extended by numerical calculations which include electron states out of the band minimum, within the $k \cdot p$ approximation  \cite{nastos2007}.  An alternate approach assumes a spherical valence band and neglects the strain-induced  modifications of the valence wavefunctions \cite{dyakonov1971b, dyakonov1974}. In all these approximate approaches, one obtains  $P_i= -0.5$.   \

In this work, we calculate using a simple analytic model the strain-induced modifications of the valence wavefunctions and we determine the oscillator strengths of the optical transitions from which we obtain $P_i^h$, $P_i^l$ and $P_i$ .  We consider uniaxial strains   along the $[0 0 1]$, $[1 1 1]$ or $[\bar1 1 0]$ crystal directions and a   light excitation  parallel or  perpendicular to the strain. Several counterintuitive results are obtained : i) The polarization does not depend on the magnitude of the strain but only on its sign. ii)  For a strain along the  $[0 0 1]$ and  $[1 1 1]$ directions, the polarization does not depend on the elasticity tensor nor on the  valence deformation potentials. The values of $P_i^h$, $P_i^l$ and $P_i$ are in agreement with the atomic model for light parallel to the strain and with the more elaborate spherical approximation for light perpendicular to the strain. iii)  In contrast, for strain along the  $[\bar1 1 0]$ direction,  $P_i^h$, $P_i^l$  depend both on the elasticity tensor and on the valence deformation potentials, a result that both the atomic model and the spherical approximation fail to predict. Finally, although the  oscillator strengths strongly depend on the directions of strain and light excitation,  the initial conduction electron spin polarization $P_i$. is always equal to -0.5 .  This suggests that the value of  $P_i$ has a universal nature. \

This paper is organized as follows. The general model for calculating the spin polarization is exposed in Sec. II, while  the atomic model and the spherical approximation are recalled  in  Sec. III. The calculation of the initial polarizations, including the strain-induced modifications of the valence wavefunctions, is shown in Sec. IV, Sec. V and Sec. VI for strain along the  [100], [111] and $[\overline 1 10]$ crystal directions, respectively. Finally, Sec. VII is devoted to the interpretation of the universal character of $P_i$.  \

\section{Model}

\subsection{Valence wavefunctions  without strain } 
In  the $\boldsymbol{xyz}$ frame of the cubic crystal, the hole  hamiltonian without strain is given by   \cite{ivchenko1997}

\begin{align}\label{Ham}
\mathscr{H}_k =  -A_k k^2 \hat{\boldsymbol{1} } +&B_k   \sum_{\alpha =x, y, z} \hat{J} _{\alpha }^2  (k_{\alpha }^2 -k^2/3) + \nonumber \\ &\frac{2}{\sqrt{3}}D_k   \sum_{\alpha =x, y, z} [\hat{J}_{\alpha }, \hat{J}_{\beta }]_+ k_{\alpha } k_{\beta} 
\end{align}
\noindent
where $ \hat{\boldsymbol{1} }$ is the unit operator,  $\boldsymbol{k}$ is the hole momentum, and $ \hat{J} _{\alpha }$ are the operators for  the hole angular momentum, for which the matrix elements on the $|3/2, \pm3/2>$  $|3/2, \pm1/2>$ basis are tabulated in Ref.   \cite{ivchenko1997}. Here $[\mathscr{K}, \mathscr{L}]_+$ is the symmetrisor of operators  $\mathscr{K}$ and $\mathscr{L}$,  and $A_k$, $B_k$, and $D_k$ are related to the usual Luttinger parameters $\gamma_1$,  $\gamma_2$  and $\gamma_3$  by  $A_k=-(h^2/2m_0)\gamma_1$, $B_k=-(h^2/m_0)\gamma_2$, and  $D_k=-(\hbar^2\sqrt{3}/m_0)\gamma_3$  where $m_0$  is the vacuum electron mass. For quantization along the $z$ direction, taken along one edge of the elementary lattice cubic cell, the hole wavefunctions are  given by

 \begin{align}\label{lh}
&|3/2, + 1/2>_z=\frac{1}{ \sqrt{6}} [ 2\:|\uparrow  >\mathscr{Z} - |\downarrow > (\mathscr{X} +i \mathscr{Y})]  \nonumber \\ & 
|3/2,- 1/2>_z=  \frac{1}{ \sqrt{6}}[  2\: |\downarrow >\mathscr{Z}+ | \uparrow> (\mathscr{X} -i\mathscr{Y})] 
\end{align}
 for the  light hole  band, and 
\begin{align}\label{hh}
&|3/2, + 3/2>_z=-|\uparrow > (\mathscr{X} +i \mathscr{Y})/ \sqrt  {2}   \nonumber \\ & 
|3/2, -3/2>_z= |\downarrow > (\mathscr{X} -i \mathscr{Y})/ \sqrt {2}  
\end{align} 
for the heavy one.  The functions $|\uparrow (\downarrow)>$ are the spin wavefunctions and the imaginary functions $\mathscr{X}$, $\mathscr{Y}$ and $\mathscr{Z}$  are a basis of the $\Gamma_5$  vector representation of the symmetry group $T_d$ and represent the degenerate hole states without spin-orbit splitting. The hole states set  defined by  Eq. \ref{lh}  and  Eq. \ref{hh} constitute a basis of the $\Gamma_8$  irreducible representation (irrep.) of the $T_d$ group. While the hole  hamiltonian  and wavefunctions are described by  Eq. \ref{Ham},  Eq. \ref{lh}  and  Eq. \ref{hh}, respectively, the corresponding valence hamiltonian is obtained by changing the sign of  Eq. \ref{Ham}  while the valence wavefunctions are obtained using the time reversal operator $\hat{K} $ due to which   $|3/2, -m>_{hole} =  \hat{K}  |3/2, m>_{valence}=  (-1)^{j-m}  |3/2, - m>^*_{valence}$.\

The conduction electron wavefunctions are of the form  $\mathscr{S} |1/2 \pm 1/2>$, where  $\mathscr{S}$ is a function transforming according to the $\Gamma_1$ scalar representation of the cubic crystal. In the vicinity of the zone  center,  the  matrix representation of the hamiltonian $ \mathscr{H}_k$  in the above standard basis takes the form  \cite{ivchenko1997}

\begin{equation}\label{Hammat}
\mathscr{H}_k= 
\begin{pmatrix}
F_k &H_k & I_k & 0 \\
H_k ^* & G_k&0&  I_k\\
I_k^* & 0 & G_k& -H_k\\
0 & I_k^* & -H_k^*& F_k
\end{pmatrix}
\end{equation}

\noindent
where 
\begin{equation}\label{Fk}
F_k (G_k)=-A_kk^2 + (-) \frac{B_k}{2}(3k_z^2-k^2)
\end{equation}
 \begin{equation}\label{Hk}
H_k =D_k k_z (k_x-ik_y)
\end{equation}
 \begin{equation}\label{Ik}
I_k =\frac{\sqrt{3}}{2}B_k  (k_x^2-k_y^2) -iD_k k_xk_y
\end{equation}

Diagonalization of   $  \mathscr{H}_k$ allows us to  find two doubly  degenerate  levels, of energies with respect to the bottom of the unperturbed hole band given by 

\begin{align}\label{En}
E_{j} = -A_{k}k^2  & + (-1)^{j +1} \Big\{ \frac{3}{2} \sum_{\alpha =x, y, z}[B_{k} (k_{\alpha }^2 -k^2/3)]^2 +  \nonumber \\ & +\sum_{\alpha >\beta}[ D_{k}k_{\alpha } k_{\beta} ]^2 \Big\}^{1/2}  
\end{align} \

\noindent
Here,  the highest  hole level, corresponding to the lowest valence level,  is reached for $j=1$ and the lowest  one for $j=2$. In the valence basis, the wave fonctions associated with the energies $- E_j $ are those of a pseudospin, and are obtained by time reversal from the hole eigenstates. They are given by
  
 \begin{align}\label{fonc1}
\psi _{j, 1}  =\frac{1}{\sqrt{(E_{j}-F)(E_j -E_i)}}
\begin{pmatrix}
I \\
 0\\
(E_j-F )\\
-H ^* 
\end{pmatrix}
\end{align}

 \begin{equation}\label{fonc2}
\psi_{j, 2}  =\frac{1}{\sqrt{(E_{j}-F )(E_j -E_i)}}
\begin{pmatrix}
H \\
 E_j-F \\
0\\
 I ^* 
\end{pmatrix}
\end{equation}
where $I= I_k$, $H= H_k$ and  $F=F_k$ and $i \neq j$. Here the phase factors are chosen such that $\psi_{i, 2}  = \hat{K}\psi_{i, 1}$ so that these states are mutually related by time reversal. \

\subsection{Effect of  a uniaxial strain on the valence wavefunctions} 
The  displacement tensor $[\boldsymbol{\epsilon}]$ induced by application of a uniaxial strain is  related to the strain   tensor $[\boldsymbol{\sigma}]$    by the general matrix relation, valid for a cubic symmetry \cite{yu1996}    
\
\begin{equation}\label{epsmat }
\begin{pmatrix}
 \epsilon_{xx}\\
 \epsilon_{yy}\\
 \epsilon_{zz}\\
2 \epsilon_{yz}\\ 
 2\epsilon_{zy}\\ 
2\epsilon_{xy}  
\end{pmatrix}
 = 
\begin{pmatrix}
S_{11}&S_{12}&S_{12}  & 0& 0  & 0\\
S_{12} &S_{11}&S_{12}& 0& 0& 0\\
S_{12} & S_{12} & S_{11}& 0& 0& 0\\
0  & 0& 0&  S_{44} &0& 0\\
0  & 0& 0& 0&  S_{44} & 0\\
0  & 0& 0& 0&   0  & S_{44}
\end{pmatrix}
\times
\begin{pmatrix}
\sigma_{xx}\\
 \sigma_{yy}\\
 \sigma_{zz}\\
 \sigma_{yz}\\ 
 \sigma_{zy}\\ 
\sigma_{xy}  
\end{pmatrix}
\end{equation}
where $[\boldsymbol{S}]$    is the elasticity tensor. Because of cubic symmetry, this tensor depends on only three elements, $S_{11}= 1.16 \times 10^{-12}$ cm$^2$/dyn, $S_{12}= -0.37 \times 10^{-12} $ cm$^2$/dyn  and $S_{44}= 1.67 \times 10^{-12} $ cm$^2$/dyn  \cite{gavini1970}.  \

Using the invariant method  \cite{bir1974},  one finds that  the valence wavefunctions are   given by  Eq.  \ref{fonc1} and  Eq.  \ref{fonc2},  replacing in Eq. \ref{Hammat} $F_k$, $H_k$  and  $I_k$  by  $F_k +F_{\epsilon}$, $H_k + H_{\epsilon}$  and  $I_k +I_{\epsilon}$, respectively  \cite{note19} and therefore  the hole hamiltonian $\mathscr{H}_k$ by  $\mathscr{H} =   \mathscr{H}_{k}+  \mathscr{H}_{\epsilon}$. \
The quantities  $F_{\epsilon}$, $G_{\epsilon}$, $H_{\epsilon}$ and $I_{\epsilon}$ are obtained by replacing  $ k^2 $ by $\epsilon =  \epsilon_{xx}+  \epsilon_{yy}+ \epsilon_{zz}$ and $k_{\alpha }k_{\beta }$ by  $\epsilon_{ \alpha \beta}$ given respectively by 
 
 \begin{equation}\label{Feps}
F_{\epsilon} (G_{\epsilon} )=-A_{\epsilon} \epsilon  + (-) \frac{B_{\epsilon} }{2}(3\epsilon _{zz}-\epsilon)
\end{equation}
 \begin{equation}\label{Heps}
H_{\epsilon}  =D_{\epsilon}  (\epsilon_{zx}-i\epsilon_{zy})
\end{equation}
 \begin{equation}\label{Heps}
I_{\epsilon}  =\frac{\sqrt{3}}{2}B_{\epsilon}  (\epsilon_{xx}-\epsilon_{yy}) -iD_\epsilon \epsilon _{xy}
\end{equation}

\noindent
where $\epsilon =   \sum  \epsilon_{ \alpha \alpha }$  \cite{note28}. For GaAs, the values of the quantities $ A_{\epsilon}$ , $ B_{\epsilon}$ and $ D_{\epsilon}$ for hole levels  have been given  as  $   \approx -  8 \pm 1.5 $ $eV,   -2\pm 0.15 $ $eV$ and  $ -5\pm0.5 $ $eV$, respectively   \cite{ivchenko1997},  \cite{note29}.\
\noindent

Diagonalization of   $  \mathscr{H}_k  + \mathscr{H}_{\epsilon}$ allows us to  find two doubly  degenerate  hole  levels. Their energies with respect to the bottom of the unperturbed hole band are given by 

\begin{align}\label{En}
E_{j} = -A_{k}k^2  & -A_{\epsilon}\epsilon+ (-1)^{j +1} \Big\{ \frac{3}{2} \sum_{\alpha =x, y, z}[B_{k} (k_{\alpha }^2 -k^2/3) +  \nonumber \\ & B_{\epsilon}   (\epsilon_{\alpha  \alpha } -\epsilon /3)]^2 +\sum_{\alpha >\beta}[ D_{k}k_{\alpha } k_{\beta}+  D_{\epsilon} \epsilon_{\alpha \beta}]^2  \Big\}^{1/2}  
\end{align} \

\noindent
Here,  the highest level is reached for $j=1$ and the lowest  one for $j=2$. \

For each of these two pseudospins, the  heavy or light valence character is determined by calculating the effective mass $m^*$, given by $1/m^*= \hbar ^{-2}\partial ^2  E_{\pm}/ \partial k^2$ in the quantization direction. Note that, at $k=0$, the components of  the vectors  appearing in  Eq. \ref{fonc1} and   Eq. \ref{fonc2} as well as the square roots are linear in $P$. Thus, the wavefunctions do not depend on the magnitude of the strain, but only on its sign $\eta$  given  by
 \begin{equation}\label{eta}
\eta   = -\frac{ P}{| P |}
\end{equation}
which is 1 (-1) for a compressive (tensile) strain.  For a compressive strain, and using now valence levels rather than hole levels,   the light  levels are above the heavy ones and therefore correspond to $j=2$   in Eq. \ref{En}. Investigation of the effect of a tensile strain will be straightforward since   $\psi_{2,1}(-\eta) = -  \psi_{1,1}(\eta)$  and  $\psi_{2,2}(-\eta) =-  \psi_{1,2}(\eta)$.\\

\subsection{Matrix elements and optical spin polarization } 
In the semi-classsical approach, the electron-photon Hamiltonian is given  to first order   by 
  \begin{equation}\label{hamiltonian}
\hat{\mathscr{H}}_{e-photon}^{\sigma^{\pm}} ( \boldsymbol{r},t) =  \frac{e}{m_0c} \boldsymbol{A_  \cdot \hat{p} }   \;  e^{-i \omega t}  +h.c.
\end{equation} 

\noindent
where $h.c.$ denotes the hermitian conjugate,  $e$ is the absolute value of the  electron charge, $c$ is the speed of light,  $ \bf{A}$ is the vector potential  of the classical electromagnetic field and $\bf{\hat{p}}$ is the linear momentum of the electron.   For  $\sigma^{\pm}$ light excitation along the $\zeta$ direction, the operator $ \bf{A_  \cdot \hat{p} }$  can be rewritten in the circular basis  $A_{+}^{*} \cdot \hat{p}_ +  + A_{-}^{*} \cdot \hat{p}_ - + A_{0}^{*} \cdot \hat{p}_0$ where we define the coordinates of $ \bf{A }$ in    the orthonormal frame  $\xi \mu \zeta $ as\  
\begin{align}\label{Ap}
&A_{\pm}=(\mp  A_{\xi}-iA_{\mu})/\sqrt{2} \nonumber \\
&A_{0}= A_{\zeta} \nonumber \\ 
&\hat{p}_{\pm}^{\zeta}=(\mp \hat{p}_{\xi}-i\hat{p}_{\mu})/\sqrt{2}\nonumber \\ 
&\hat{p}_{0}^{\zeta}= \hat{p}_{\zeta}
\end{align}\

\noindent
The matrix elements of this  hamiltonian between conduction wavefunctions   $\Psi_{\boldsymbol{k}, c, s}= \exp [i \boldsymbol{k \cdot}r] \psi _{c, s}$  and  valence  wavefunctions $ \Psi_{\boldsymbol{k}, j,\tau}= \exp [i \boldsymbol{k \cdot}r] \psi _{j, \tau}$ are  then

\begin{align}\label{matelbis}
< \Psi _{\boldsymbol{k}, c, s}|& \hat{\mathscr{H}}_{e-photon}( \boldsymbol{r}, t) | \Psi _{\boldsymbol{k}, j,\tau}> \approx  \frac{eA_0}{2m_0c} \nonumber \\ & \left[<  - e^{-i\omega t} \mathscr{A}_{\pm}^{\zeta} (s, j, \tau)(\boldsymbol{k})  + e^{i\omega t} \mathscr{A}_{\mp}^{\zeta} (s, j,  \tau)(\boldsymbol{k})   \right]
\end{align}
where   
\begin{equation}\label{matel}
\mathscr{A}^{\zeta}_{\pm} (s, j, \tau)(\boldsymbol{k})= < \Psi _{\boldsymbol{k}, c, s}| p_{\pm}^{\zeta}| \Psi _{\boldsymbol{k}, j, \tau}>
\end{equation}
 
\noindent
 These matrix elements,  given in the supplementary material,  are obtained from those of $p_{\alpha}$, where $\alpha$ is    $\xi$,  $\mu $  or  $\zeta$ \cite{note24}.  Here, it is assumed that the carrier kinetic energy is weak, so that $\mathscr{A}^{\zeta}_{\pm} (s, j, \tau)(\boldsymbol{k}) \approx \mathscr{A}^{\zeta}_{\pm} (s, j, \tau)(\boldsymbol{0}) $. Using first order time-dependent perturbation theory and after integration over linear momentum $\boldsymbol{k}$, the state of the electron-hole system at a time $t$ after the photogeneration,  larger than the inter-band dephasing but  shorter  than the intra-band relaxation times \cite{oudar1984, kaindl2001},  is described by the density matrix 
 \

 \begin{align}
\label{psi}
 & \hat{\rho}^{\sigma^{\pm}}(t) = | \oslash  > < \oslash|+   \frac{2 \pi t  C ^2 }{\hbar } \sum_{j=1}^{2}   \mathscr{D}_j(\hbar \omega) \! \nonumber \\ & \sum_{\substack{ s, s'=\pm 1/2 \\ \tau, \tau'= 1, 2\\  }}\!   (-1)^{\tau + \tau'} \mathscr{A} ^{\zeta} _{\pm} (s; j, \tau) 
  \mathscr{A} ^{\zeta*}_{\pm} (s'; j, \tau') \nonumber \\ &  | 1/2, s  > \left[  \hat{K} |\psi_{j, \tau}> \right]         \left[ <\psi_{j, \tau'}| \hat{K}^+  \right]  <1/2, s'|
\end{align}
\noindent
where $ | \oslash >$  is the crystal ground state,  $ \mathscr{D}_i(\hbar \omega) $ is the joint density of states from the valence band $j$ to the conduction one at the excitation energy  $\hbar \omega$, and $C=(eA_0)/(2\hbar m_0c)$  \cite{note27}. The generation matrix of electron-hole pairs per unit time is thus 
 
 \begin{align}
\label{drodt}
 &  \frac{\partial \hat{\rho}}{\partial t}\approx  \frac{2 \pi  C ^2}{\hbar }  \sum_{j=1}^{2}   \mathscr{D}_j(\hbar \omega) \nonumber \\ &\sum_{\substack{ s, s'=\pm 1/2 \\ \tau, \tau'= 1, 2\\  }}  (-1)^{\tau +  \tau '} \mathscr{A} ^{\zeta}_{\pm} (s, j, \tau) 
  \mathscr{A} ^{\zeta*}_{\pm} (s', j, \tau')    \nonumber \\ &  | 1/2, s  > \left[  \hat{K} |\Psi_{j, \tau}> \right]         \left[ <\Psi_{j, \tau'}| \hat{K}^+  \right]  <1/2, s'|
\end{align}

After taking the trace over the hole states, one obtains the generation matrix of conduction electrons photogenerated from a specific valence band $j$. 
 
 \begin{align}\label{matdens }
& \frac{\partial \hat{\rho}_{c, j}}{\partial t}  =  \frac{2\pi}{\hbar}( \frac{e}{m_0 c} \frac{A_0}{2} )^2 \mathscr{D}_j(\hbar \omega)  \sum_ {\tau=1,2}\nonumber \\ & 
\left(\begin{smallmatrix}
| \mathscr{A}_{\pm}^{\zeta}  (+1/2, j, \tau)|^2 &  \mathscr{A}_{\pm}^{\zeta}  (+1/2, j, \tau) \mathscr{A}^{\zeta *}_{\pm}  (-1/2, j, \tau)  \\
 \mathscr{A}_{\pm} ^{\zeta *}(+1/2, j, \tau) \mathscr{A}_{\pm}^{\zeta}  (-1/2, j, \tau)   & | \mathscr{A}_{\pm}  ^{\zeta}  (-1/2, j, \tau)|^2    \\
\end{smallmatrix} \right)
\end{align}
The average electron spin polarization vector before spin relaxation for transitions from the valence band $j$ is  $  \propto Tr ( \hat{\rho}_{c, j} \hat{ \sigma}) $, where  $\hat{\sigma}$ are the  Pauli  matrices along the $\xi$,  $\mu $  or  $\zeta$ directions,  normalized by the pair generation rate  ($  \propto Tr (\hat{\rho}_{c, j}) $). The early time dependence is eliminated in the normalization. For a $\sigma^+$ light excitation, one obtains  at  time delays smaller than  spin relaxation characteristic times \

 \begin{align}\label{spin}
& \boldsymbol{ \hat{S}}_j ^{\zeta} = \frac{\hbar}{2} \frac{1}{\sum_{\substack{ s, s'=\pm 1/2 \\ \tau = 1, 2\\  }} | \mathscr{A}^{\zeta} (s, j, \tau)|^2}  \sum_ {\tau=1,2}  \nonumber \\ & 
\left(\begin{matrix}
 2Re[\mathscr{A}_+^{\zeta*}  (1/2, j, \tau) \mathscr{A}^{\zeta}_+  (-1/2, j, \tau) ]  \\
 2Im [\mathscr{A}^{\zeta*}_+  (1/2, j, \tau) \mathscr{A}^{\zeta}_+  (-1/2, j, \tau) ]\\
 |  \mathscr{A}^{\zeta}_+  (1/2, j, \tau)|^  2 - |  \mathscr{A}^{\zeta}_+  (-1/2, j, \tau)|^  2
\end{matrix} \right)
\end{align}

The   overall spin, including all transitions but neglecting the effective mass correction,  is given by  

\begin{equation}\label{polartot 110y'}
\boldsymbol{ \hat{S}}^{\zeta} = \frac{ \boldsymbol{ \hat{S}}_h ^{\zeta} + r_{l,h}   \boldsymbol{ \hat{S}}_l ^{\zeta}}{1+r_{l,h} }
\end{equation}
where $r_{l,h}=r_{l}/r_{h}$  is the population ratio between photogenerated light- and heavy-holes, defined within a multiplicative constant by:
  
\begin{equation}\label{rerh}
r_{j}=  Tr( \hat{\rho}_{c,j }) 
\end{equation}
 In the general case, there exists more than one possible transition towards a given conduction state,  so that  the above density matrix   may have nondiagonal elements implying that the mean electronic spin may lie out of the $\zeta$ direction. \

It is shown in Appendix A  that  a change of excitation light helicity  results in   an interchange of the terms corresponding to $\tau=1$ with those corresponding to $\tau=2$  with a possible change of sign. As a consequence,  a change of light helicity results in a change of sign of the spin -polarization. In the following, we therefore only consider excitation by a $\sigma ^+$ light excitation. \

\section{Background}

In this section, for comparison purposes, we explain the two simplifying  models which have been used in the past to calculate the photoelectron spin polarization. \

\subsection{The atomic model}

In   the case  of a zero strain and assuming that the electron and hole momenta lie along the direction of light excitation, along which the spins are quantized, the  conduction electron states  belong to the $\Gamma _6$ representation of $T_d$ for which a basis is $ |\mathscr{S}> |1/2, \pm 1/2>$, where the orbital part $ |\mathscr{S}>$  belongs to the scalar representation  $\Gamma _1$.  The valence levels, described by   Eq. \ref{lh} and  Eq. \ref{hh}, are a basis of the $\Gamma_8$ representation.   The vector potential $ \bf{A}$ and  the operator  $\bf{\hat{p}}$ both belong to the $\Gamma_5$ representation  of $T_d$. It can be seen from the coupling tables  \cite{koster1963}  that the only significant matrix elements of the electron impulsion vector-operator $\bf{\hat{p}}$   are   \cite{note20} 
\begin{equation}\label{picv}
\Pi_{cv} =<\mathscr{S}|\hat{p_x} |\mathscr{X}>=   <\mathscr{S}|\hat{p_y}|\mathscr{Y}>=   <\mathscr{S}|\hat{p_z}|\mathscr{Z}>
\end{equation}

\begin{figure}[tbp]
 \includegraphics[clip,width=8cm] {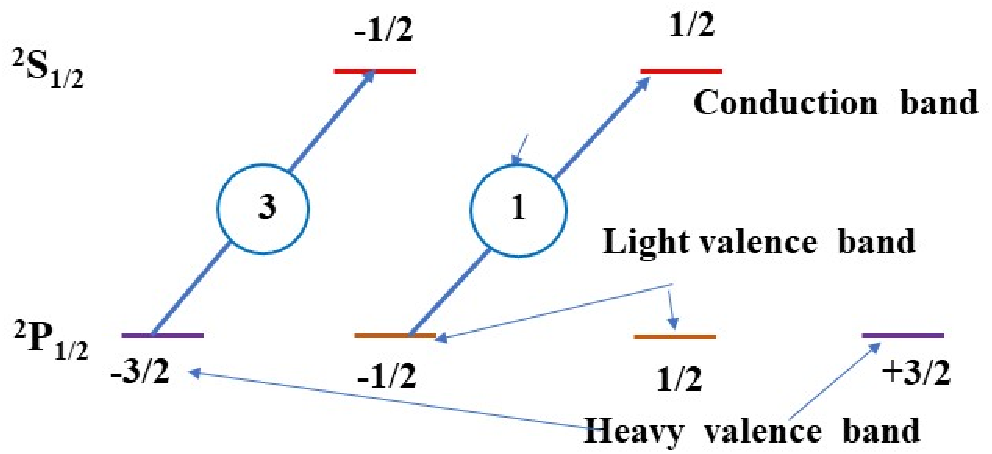}
\caption{ Relative oscillator strengths for  $\sigma^{+}$ light excitation within the  usually accepted atomic  model, under zero strain. This picture implies that the photon orbital momentum is transferred to the electron so that, under light absorption,  the quantum number $m$ changes by  $+1$.  The initial polarization for transitions from the heavy (light) valence levels is $-1$ ($+1$), leading to an overall spin polarization of $-1/2$. This  picture is valid only if the direction of the electron momentum $\boldsymbol{k}$ coincides with that of light excitation.}
\label{Fig1}
\end{figure}

\noindent
where, with our phase conventions, $\Pi_{cv}$ is a real quantity. Thus,  as shown in Fig. \ref{Fig1},  the only possible transitions under $\sigma^+$- polarized light excitation are from the $ |3/2, - 3/2>$ heavy valence level to the  $ | 1/2 - 1/2>$  conduction level, and from the $ |3/2, - 1/2>$ light valence level to the  $ |1/2 +1/2>$  conduction level. There is only one possible transition leading to a given conduction electron spin state. Thus, each matrix  labelled by $j$ in Eq. \ref{spin}  has only one element, on the diagonal, so that the polarization lies  along the quantization axis  which is also the propagation direction of light. Here the initial electron polarization from the heavy  (light) valence levels  is   $  \mp 1$  ($  \pm 1$)  for a  $\sigma ^{\pm}$-polarized light. Taking into account the oscillator strengths, the overall conduction spin polarization  has the well-known value $P_i = \mp 1/2$.  \

Within this simple model,  absorption of $\sigma ^{\pm}$  light can occur provided the change in angular momentum is $\Delta m= \pm1$, which suggests  that the angular momentum of the photon is transferred to the photocreated electron-hole pair  \cite{dyakonov1971b}..  \

\subsection{Spin orientation within the spherical approximation}
Here, we consider the  approximation  where i)  the valence levels are not affected by the applied strain and where ii) the spatial anisotropy of the valence band (warping) is neglected. The spherical approximation would be strictly valid if $B_k = D_k/\sqrt(3)$, and in the strained case, if $B_{\epsilon}= D_{\epsilon}/\sqrt(3)$ and $S_{11} -S{12}=S{11}/2$, so that the effective hamiltonians $\mathscr{H}_k$ and $\mathscr{H}_{\epsilon}$  acquire a  spherical symmetry. However, this is only roughly approximate in Td crystals, as can be checked from experimental data.\
 
In this picture, the valence bands  can be quantized either along the direction of the light propagation $\boldsymbol{q}$, which can be labelled as the z axis, without strain, or along the strain direction $\boldsymbol{u}$. The valence states removed by photo-generation are conveniently expressed taking their wave vector $\boldsymbol{k}$ as their quantization axis. They are then expressed as linear combinations of  the   valence states quantized along $\boldsymbol{q}$, $|3/2,m>|_{\boldsymbol{q}}$  or along  $\boldsymbol{u}$ (see details of the calculations  in the supplementary material).\
 
In the absence of any strain,  after integration over the azimuthal angle $\varphi$, the dependences as a function of polar angle $\theta$  of the concentration and spin polarization of electrons photo-generated from the heavy-valence band   are,  respectively

 \begin{align}\label{depthetahh}
  &n_{hh}(\theta)  \propto   1+cos^2\theta  \nonumber \\ &
 \mathscr{P}_{hh}(\theta) =-2  \frac{cos^2\theta }{1+cos^2\theta }
\end{align}
\noindent
while the corresponding values for transitions from light valence levels are given by
 \begin{align}\label{depthetalh}
  &n_{lh}(\theta)  \propto   \frac{2}{3} +sin^2\theta  \nonumber \\ &
   \mathscr{P}_{lh}(\theta) = \frac{1-3sin^2\theta }{1+\frac{3}{2} sin^2\theta}
\end{align} 

 If the polar angle $\theta$, is zero, in agreement with the atomic model, the initial polarizations are equal to -1 and +1  for transitions from heavy or light valence levels.   For $0<\theta <\pi $, both the conduction electron concentrations and spin polarization are  modified revealing a strong effect of the direction of $\boldsymbol{k}$ on the optical orientation and the fact that the atomic  model is no longer valid. The strongest effect is observed for $\theta= \pi/2$, since for this value $\mathscr{P}_{hh}=0$, while  $\mathscr{P}_{lh}=-4/5 $  has a sign opposite to the one at   $\theta= 0$. Such finding reveals a type of spin-momentum coupling process under light absorption. \

Conservation of angular momentum, implied by the atomic model,  fails as soon as $\boldsymbol{k}$  does not lie along $z$. Although this conservation  is still valid for the valence wavefunctions  $|\frac{3}{2}, m>_{\boldsymbol{q}}$  quantized along $\boldsymbol{q}$, it is no longer true for the  wavefunctions quantized along  $\boldsymbol{k }$, since these functions  are linear combinations of the $|\frac{3}{2}, m'>_{\boldsymbol{q}}$  ($m'=\pm 3/2,  \pm 1/2)$. \

The spin polarization of conduction electrons induced by $\sigma^+$  light excitation is finally obtained by integration of the generation  matrices over $\theta$. For transitions from heavy and light valence levels, the corresponding values do not depend on the joint density of states functions and are  given  by $<  \mathscr{P}_{hh}>= -1/2$ and  $<  \mathscr{P}_{lh}>=- 1/2$, respectively. This implies that,  in agreement with the atomic model,  the average conduction electron spin polarization is equal to $-1/2$. . \ 

%\begin{figure}[tbp]
% \includegraphics[clip,width=6cm] {Figsphv13.eps}
%\caption{The inset  defines the  polar angles ($\theta$, $\phi$) characterizing the generated electron wave vector. The main figure shows the dependence as a function of $\theta $ of conduction electron concentration and spin polarization generated from heavy and light valence levels within the spherical approximation and in the absence of a strain  [Eq. \ref{depthetahh} and   Eq. \ref{depthetalh}].}
%\label{Fig03}
%\end{figure}

%\subsection{Optical selection rules}
 
%\subsection{Optical orientation under strain in the spherical approximation  } 

For a strained semiconductor, as shown in the supplementary material, the mean electronic spins for transitions from heavy and light valence levels are finally given by

 \begin{align}\label{depstrain} 
 &\vec{\mathscr{S}} _{hh}  =-\frac{\hbar}{2}  \frac{2(\vec{u} \cdot  \vec{e}_q) \vec{u} }{1+(\vec{u} \cdot  \vec{e}_q)^2 }  \nonumber \\ &
 \vec{\mathscr{S}} _{lh}  = \frac{\hbar}{2}2 \frac{3(\vec{u} \cdot  \vec{e}_q) \vec{u} -2 \vec{e} _q}{5-3(\vec{u} \cdot  \vec{e}_q)^2 }
\end{align}
These equations imply that $ \vec{\mathscr{S}} _{hh}$ and $\vec{\mathscr{S}} _{lh}$ depend only on the angle between the light and strain directions. In the general case, the spin polarization generated from the heavy valence levels lies along  the direction of the strain, while the spin polarization generated from the light valence levels lies in the   $(\vec{e}_q,  \vec{u})$ plane. \

In the limit case where the light excitation is parallel to the strain direction, the spin polarization lies along this common direction and is equal to $-1$ for excitation from  the heavy valence levels and to $+1$ for excitation from the light valence levels. This is identical with the results obtained by the atomic model. Conversely,  and in strong contrast with the atomic model,  if the light excitation is perpendicular   to the strain direction,  the  spin polarization generated from heavy  valence levels is zero, while that generated from light valence levels,  lying along the direction of light excitation, is negative and equal to $-4/5(\hbar /2)$. \

\section{Strain along [0 0 1] }

%\subsection{Heavy-light valence splitting for a strain along [001] }
The case of a uniaxial pressure  $P$ applied along the z direction is found in situations where an epitaxial layer grown on a (0 0 1) substrate has a lattice mismatch with its substrate  \cite{egorov2005, tessler1994}. In this case the resulting biaxial strain in the plane of the substrate is equivalent to an opposite uniaxial strain in the perpendicular, [0 0 1] direction. The strain tensor has only one nonzero component $\sigma _{zz}= P$  and the non-zero components of the displacement are $\epsilon_{x x} =\epsilon_{y y} = S_{1,2} P$ and $ \epsilon_{z z} = S_{1,1} P $.  The terms $H_{\epsilon}$ and $I_{\epsilon}$  are zero so that the only nonzero components of  $\mathscr{H}_{\epsilon}$ are 

\begin{figure}[tbp]
 \includegraphics[clip,width=8 cm] {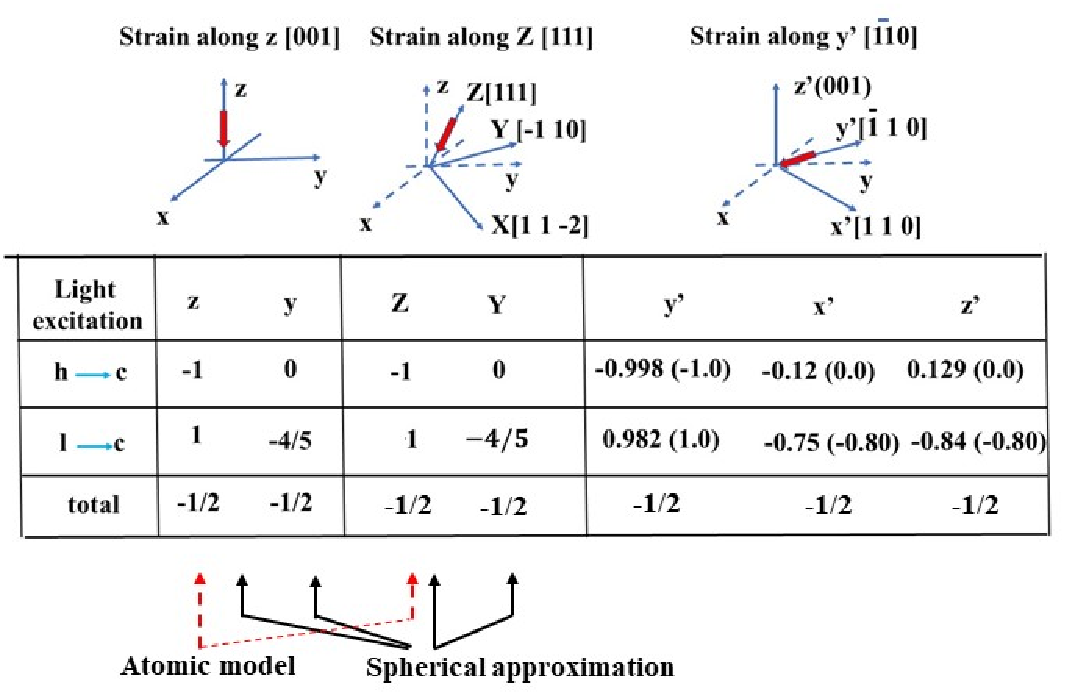}
\caption{The top panel shows the various orientations of the strain considered here and defines the corresponding  axis labels. The table shows, for each strain direction,  the calculated spin polarization of   conduction electrons generated by $\sigma ^+$-polarized light excitation  of orientation parallel (left column) and perpendicular to the strain direction. In all these cases, the conduction electron spin polarization lies along the direction of light excitation and the overall spin polarization including all transitions is $-0.5$.   When the light excitation lies along the strain direction, the above values coincide with the predictions of the atomic model [see Fig. \ref{Fig10}]. If  the light excitation is perpendicular to the strain direction,  the values  are in agreement with the spherical approximation [Eq. \ref{depstrain}], except for a  strain along the $|\bar 1 1 0]$   orientation, where  the polarization depends on the elasticity and deformation potential tensors. In this case, the values in parentheses give another set  of possible values within the uncertainties given in  Ref. \cite{ivchenko1997}.  The arrows show the cases where, respectively,  the atomic model and the spherical approximation are valid.   }
\label{Fig2}
\end{figure}
\begin{align}\label{H100}
 & \mathscr{H}_{\epsilon11}=  \mathscr{H}_{\epsilon44}=-A_{\epsilon}(S_{11} +2S_{12})P+\ B_{\epsilon} (S_{11} -S_{12})| P \nonumber \\ &  \mathscr{H}_{\epsilon22}=  \mathscr{H}_{\epsilon33}=-A_{\epsilon}(S_{11} +2S_{12})P-\ B_{\epsilon} (S_{11} -S_{12}) P 
\end{align} \
This matrix is diagonal and, for  $\boldsymbol{k}= k [0 0 1]$, the hole eigenenergies are simply given by 
 
\begin{align}\label{Enk100}
  E_{\pm} \    = -A_{k}k^2 &-A_{\epsilon}(S_{11} +2S_{12})P \nonumber \\ &  \pm [|B_{k} k^2 +  B_{\epsilon}   (S_{11} - S_{12})P]
\end{align} \\
 
\noindent
and the corresponding hole effective masses at the center of the Brillouin zone in the  [0 0 1] direction are given by

 \begin{equation}\label{mh100}
\frac{1}{m^*_{h\pm}}= \frac{1}{m_0}\Big\{\gamma_1 \mp 2  \eta \gamma_2\Big\}
\end{equation} 
 
\noindent
Since the splitting between the heavy and light valence levels is smaller than the valence spin-orbit parameter, the coupling of the light-hole band with the split-off valence state and with the upper conduction bands can be neglected  \cite{bertho1994},  \cite{note20, note22}. The $T_d$ conduction spin states of the irrep. $\Gamma_6$ are compatible with those of $D_{2d}$, as can be checked from the compatibility tables \cite{koster1963}. The  $\Gamma_8$ valence states in turn split into two two-dimensional irrep.,  $\Gamma_6$  for the heavy valence states and  $\Gamma_7$  for the light valence ones. The zero strain wavefunctions, defined in Eq. \ref{lh} and Eq. \ref{hh}, are compatible and can be taken as basis in their respective irrep.  In the case of  light excitation along the direction z  of the strain,  and as indicated in   Fig. \ref{Fig2}, the results correspond to the atomic model as well as to  the spherical approximation. The oscillator strengths  are the same as given in Fig. \ref{Fig1}. \

\begin{figure}[tbp]
\includegraphics[clip,width=8 cm] {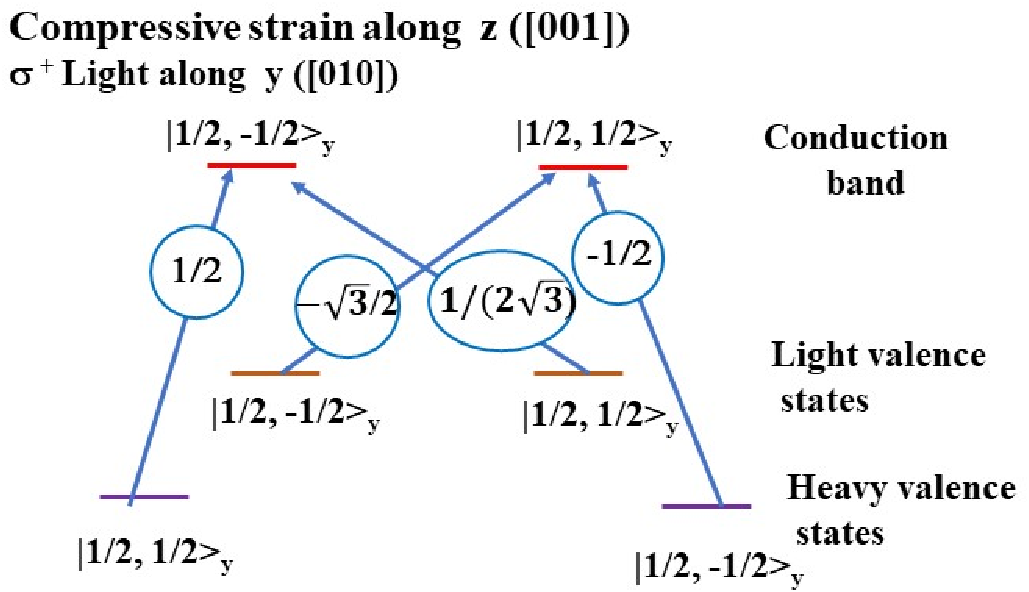}
\caption{Matrix elements for the optical transitions for a compressive strain under the [001]  direction  and a light excitation perpendicular to the strain. These elements are expressed in units of $ \Pi_{cv} $, given by Eq. \ref{picv} and the valence states are represented by pseudo-spins along the $y$ quantization axis.. The oscillator strengths of the transitions are proportionnal to the squared modulus of these matrix elements. Transitions from the light  valence band generate an initial  spin polarization of conduction electrons of $-4/5$, while transitions from the heavy valence band do not generate any spin polarization. }
\label{Fig3}
\end{figure}

For a light excitation propagating perpendicularly to the strain direction, the conduction and valence pseudo-spins must be quantized along this direction of light excitation. For the $\boldsymbol{y}$ direction, for instance, the calculation of the matrix elements can be done directly using the conduction and valence wave functions given in the supplementary material. The oscillator strengths are given in Fig. \ref{Fig3}.  It is then predicted that the conduction electron spin polarization for transitions from heavy valence band alone is zero, while that from light valence one is -4/5. The same is expected for light excitation along  the $\boldsymbol{x}$ direction.  As indicated in Fig. \ref{Fig2},  these results are in agreement with the predictions of the spherical approximation. \
\

\section{Strain along [1 1 1] }

%\subsection{Heavy-light valence splitting}

For a uniaxial strain  along the Z  direction ([1 1 1]),  all the components of the strain tensor are equal. Thus $\sigma_{\alpha \beta } = P/3$, for $\alpha, \beta = x,y,z$. The displacement tensor is given by 

\begin{equation}\label{epsilon111}
[\epsilon]= \frac{P}{6} 
\begin{pmatrix}
2(S_{11} +2 S_{12}) &S_{44}/2 &S_{44}/2  \\
S_{44}/2 & 2(S_{11} +2 S_{12})&S_{44}/2 \\
S_{44}/2  & S_{44}/2 & 2(S_{11} +2 S_{12}) \\
\end{pmatrix}
\end{equation}
 
\noindent
The  effective Hamiltonian of heavy and light hole states in the canonical basis, defined formally as in  Eq. \ref{Ham} with valence orbital functions,  takes the form 
\begin{align}\label{Ham2}
\mathscr{H} &_{\epsilon} = -A_{\epsilon}  \epsilon \hat{\textbf{1} } +\frac{2 D _{\epsilon}}{\sqrt{3}}   \sum_{\substack{ \alpha, \beta =x, y, z \\ \beta>\alpha }}[J_{\alpha},  J_{\beta}]_+ \epsilon_{\alpha, \beta} \nonumber \\ &
=\frac{P}{3}   \left[-3 A _{\epsilon} (S_{11} +2S_{12} ) \hat{\textbf{1}} +  \frac{D _{\epsilon}}{\sqrt{3}} S_{44}  \sum_{\substack{ \alpha, \beta =x, y, z \\ \beta>\alpha }}[J_{\alpha},  J_{\beta}]_+   \right]
\end{align}

After removing the first term by an energy translation, we see that this hamiltonian  is proportionnal to the sole coefficient $ D_{\epsilon}S_{44}/\sqrt{3}$, which implies that the normalized eigenstates do not depend on the strain tensor coefficients nor on  the deformation potentials. They only reflect the trigonal symmetry of the new eigen basis and are adapted to the representations of $C_{3v}$,  $\Gamma_4$ irrep  for light holes  and the reducible representation $\Gamma_5 +\Gamma_6$ for heavy holes.   If $\boldsymbol{k}$ is along the [111] direction, Eq. \ref{En} becomes 
 
\begin{align}\label{En111}
E_{\pm} = -A_{k}k^2  & -A_{\epsilon}(S_{11} +2S_{12})P  \pm \frac{1}{\sqrt{3}} | D_{k}k^2 +  D_{\epsilon}\frac{S_{44}}{2} P  |
\end{align} 
 The  corresponding effective masses along [1 1 1] are given by 
 
  \begin{equation}\label{mh111}
\frac{1}{m^*_{h\pm}}= \frac{1}{m_0}\Big\{\gamma_1 \mp2 \eta \gamma_3\Big\}
\end{equation} \ 

\begin{figure}[tbp]
\includegraphics[clip,width=8 cm] {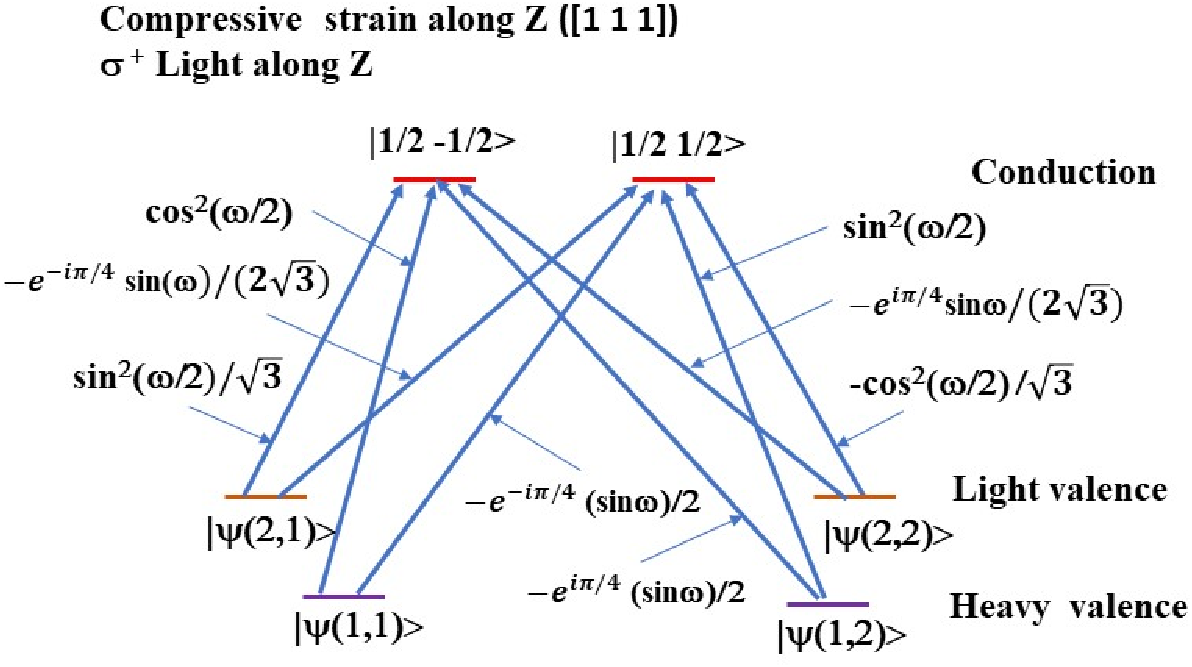}
\caption{For a compressive  strain along the Z direction, the light valence levels lie  above the heavy  valence ones. For $\sigma^+$-polarized light  along the direction of the strain, the transition matrix elements between  valence and conduction states are expressed in units of $ \Pi_{cv} $, given by Eq. \ref{picv}. Here, the angle $\omega= \boldsymbol{(z,Z)}$, between the strain direction and the crystal $z$ axis,  is given by  $cos(\omega) =1/\sqrt{3}$ and   is the value of the polar angle in the present case. The spin quantization axis is $z$.}
\label{Fig4}
\end{figure}

For a compressive strain,  (i.e. $\eta=+1$), as illustrated in Fig. \ref{Fig4}  for $k=0$, the light hole  level corrresponds to the negative  sign in  Eq. \ref{En111} and to the positive one in Eq. \ref{mh111} and lies  above  the heavy one.  The heavy-light valence splitting is given by 
 
  \begin{equation}\label{Delta111}
\Delta E = E_{+}-  E_{-} = \frac{1}{\sqrt{3}} | PD_{\epsilon}| S_{44}  
\end{equation} \

\noindent
In the valence canonical basis, the normalized eigenstates,  obtained by time reversing the hole eigenstates,  are given by

\begin{align}\label{wavefunction111}
\begin{pmatrix}
 |\psi_{1,1}>\\
|\psi_{1,2}>\\
 |\psi_{2,1}>\\
|\psi_{2,2}>\\  
\end{pmatrix}
 = 
 & \begin{pmatrix}
-\frac{\eta i}{\sqrt{6}}&0 & \frac{1 }{\sqrt{2}}&- \eta \frac{ e^{i\pi/4}}{\sqrt{3}}\\
\eta  \frac{e^{-i\pi/4}}{\sqrt{3}}& \frac{1}{\sqrt{2}}& 0& \frac{\eta i }{\sqrt{6}}\\
-\frac{\eta i}{\sqrt{6}}&0  &- \frac{1 }{\sqrt{2}} &- \eta  \frac{e^{i\pi/4}}{\sqrt{3}}\\
\eta  \frac{e^{-i\pi/4}}{\sqrt{3}}  & - \frac{1}{\sqrt{2}} & 0&\frac{\eta i }{\sqrt{6}}\\
\end{pmatrix}
 \nonumber \\ &\times
\begin{pmatrix}
|+3/2>\\
 |+1/2>\\
| -1/2>\\
 |-3/2>\\ 
\end{pmatrix}
\end{align}
 
These wavefunctions do not depend on the elasticity tensor nor on the valence deformation potentials. This is due to the fact that, as seen from Eq. \ref{epsilon111}, $2 \epsilon_{zz} -\epsilon_{xx} -\epsilon_{yy} =\sqrt{3}[\epsilon_{xx} -\epsilon_{yy}]=0$, so that the components of the elasticity tensor transforming like the   $\Gamma_3$ irrep of the $T_d$ group are zero. \
 
Using the procedure described in Sec. II, we first obtain the matrix elements  of $\hat{p}_X$,  $\hat{p}_Y$ and   $\hat{p}_Z$  between valence and conduction wavefunctions.  These elements  are given in the supplementary material. Considering for specificity that the stress is compressive,  and as shown before, the top lying valence levels  have a light character.\

 For circularly-polarized light propagating along the Z direction,  the matrix elements for absorption of circularly-polarized light are shown in Fig. \ref{Fig4}. They are conveniently expressed as a function of the angle $\omega$ such that $\cos(\omega) =1/\sqrt{3}$.  This angle is the value of the polar angle for the present orientation of  the uniaxial strain.\

It is apparent that, in this case and in  contrast with the cubic case, all optical transitions are allowed although with distinct relative probabilities. In the same way as for the cubic case, one finds that, for transitions from  the heavy valence levels induced by $\sigma^{+}$-polarized light, the conduction electron mean spin lies along the  direction $Z$ of strain and of light excitation, with a polarization $P_{e,hv}^{ +} =- 1$.  Applying the same method to the transitions from the light valence level gives the same orientation of the conduction electron spin  with an opposite polarization $P_{e,lv}^{+} = + 1$.  The overall initial spin polarization, including all transitions from valence levels, takes into account the fact that as seen in Fig. \ref{Fig4}, the  strengths of the optical transitions from heavy valence levels are a factor of 3 larger than those for light valence levels.   This gives an initial spin polarization of  $- 0.5$,   i. e.  identical to the one found for the cubic case , and to the predictions of the spherical approsimation.  Again, although the oscillator strengths are different from the predictions of the  atomic model, the predicted spin polarization values are identical. \

 \begin{figure}[tbp]
 \includegraphics[clip,width=8 cm] {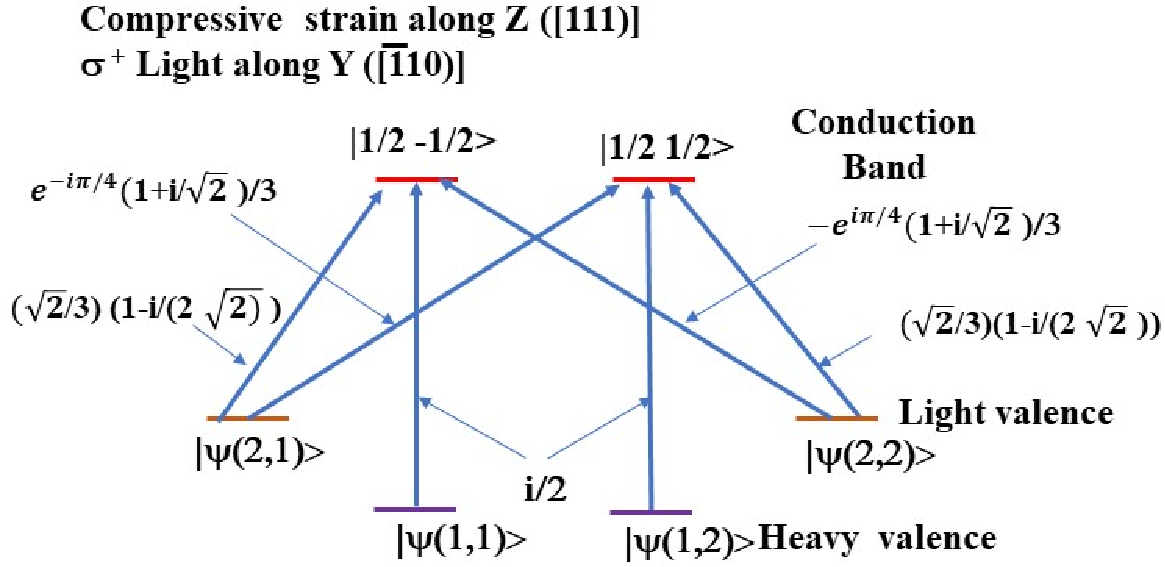}
\caption{Same as Fig. \ref{Fig4}, but for a light excitation along the $Y$ direction i. e. perpendicular to the direction of the strain.  }
\label{Fig5}
\end{figure} 

For light propagating along the $Y$ direction, as seen from  the supplementary material, for $\eta = 1$, the matrix elements of $p_Z$ for transitions from  the heavy valence levels $\psi _{1, 1}$  and $\psi _{1, 2}$ are zero, implying that the component $A_Z$, or the electromagnetic electric field component along $\boldsymbol{Z}$ is optically inactive \cite{note23}.  This implies that, for excitation  along the $Y$ direction and more generally for all modes which propagate in the X,Y plane, the electron spin  polarization generated by these transitions is zero.\ 

For  transitions from light valence states,  the matrix elements are given by  Fig. \ref{Fig5}. Again, using the generation matrix formalism,  one  finds that  the mean  electron spin is given by. \

\begin{equation}\label{spinY}
 <\boldsymbol{S}> = -\frac{\hbar}{2}  \frac{4}{5}  \boldsymbol{Y}
\end{equation}

 Finally, including all transitions from the valence levels, one finds that the polarization value is $-1/2$,  in agreement with the predictions of the spherical model.  This is in fact expected, since the valence states are independent of the elasticity and deformation tensors. Hence the optical valence-conduction matrix elements neither depend on the ratio $ B_{\epsilon}/D_{\epsilon}$  nor on $ (S_{11}-S_{12})/S_{44}$. Thus the polarization results must coïncide with those obtained in the spherical case. \\

 \section{Strain along $[\bar 1 1 0]$ }

 %\subsection{Heavy-light valence splitting}
As illustrated for  nanowires  \cite{paget2022}, a biaxial  strain in the $x'z'$ plane is equivalent to an opposite  uniaxial strain in the $y'$  ($[\bar 1 1 0]$) direction. This situation can also be experimentally achieved for an epitaxial layer on a $[100]$ substrate and a strain applied on the lateral cleavage faces, oriented along the $[110]$  and $[1\bar 1 0]$ crystallographic planes.  In the $xyz$ frame,  the displacement tensor is given by 

\begin{equation}\label{epsilon110}
[\epsilon]= \frac{P}{2} 
\begin{pmatrix}
(S_{11} + S_{12}) &-\frac{  S_{44}}{2}   &0\\
-\frac{  S_{44}}{2}  & (S_{11} +S_{12})&0 \\
0  & 0  & 2 S_{12}  \\
\end{pmatrix}
\end{equation}
  The dispersion curve for holes [Eq. \ref{En}] becomes

\begin{align}\label{En110}
&E_{\pm} =  -A_{k}k^2   -A_{\epsilon}(S_{11} +2S_{12}) P  \nonumber \\ & \pm \frac{1}{2} \Big\{ [B_{k}k^2 +  B_{\epsilon}P (S_{11} -S_{12})]^2   + [D_{k}k^2+ \frac{D_{\epsilon}  S_{44}}{2} P]^2   \Big\}^{1/2}
\end{align} 

 \noindent
Since $A_{\epsilon} < 0$ , for $k=0$,  \ Eq. \ref{En110} is reduced to 
 
 \begin{align}\label{En-110}
 E_{\pm}&  =  -A_{\epsilon}(S_{11}+2S_{12}) P \nonumber \\ & \Big\{1     \pm \eta \frac{[B_{\epsilon}^2 (S_{11}-S_{12})]^2  +\frac{ D_{\epsilon}^2  S_{44}^2}{4}]^{1/2} }{-  2A_{\epsilon}(S_{11}+2S_{12}) } \Big\}  
\end{align}

The  corresponding effective masses along $[\bar 1 1 0]$ are given by 
 
 \begin{equation}\label{mh110}
\frac{1}{m^*_{h\pm}}= \frac{1}{m_0}\Big\{\gamma_1 \mp  2 \eta \frac{\gamma_2 | B_{\epsilon}| (S_{11} -S_{12})+ \gamma_3 |D_{\epsilon}|S_{44}/2}{[B_{\epsilon}^2(S_{11} -S_{12})^2 +D_{\epsilon}^2\frac{  S_{44}^2}{4}]^{1/2}}\Big\}
\end{equation}
\noindent  
and are modified by application of the stress.   These results are in agreement with symmetry arguments. Using the compatibility table between $T_d$ and $C_{2v}$ \cite{koster1963}, one sees that both the heavy and light valence states belong to the  $\Gamma_5$ two-dimensional  spinor representation of  $C_{2v}$. \\

Since the hole levels satisfy $E_+>E_-$ and since  $S_{11} -S_{12} >0$ and  $S_{44}>0$,  for a compressive uniaxial stress   along $[\bar 1 1 0]$,  such that  $\eta= 1$,   the lower hole (higher valence)  level, denominated $\phi_{2, \tau}$  and satisfying $|\phi_{2, \tau}>= \hat{K} |\psi_{2, \tau} >$,   corresponds to the light mass, while    the upper  hole (lower  valence)  level, denominated $\phi_{1, \tau}$  and satisfying $|\phi_{1, \tau}>= \hat{K} |\psi_{1, \tau} >$,   has a heavy valence character.  Conversely, for a tensile strain,    the lower hole  (higher valence) level has a heavy character, while the upper hole (lower  valence)  level has a light character.\ 

In the canonical valence basis defined by  Eq. \ref{lh}  and Eq. \ref{hh}, the valence wavefunctions depend on the sign of $P$  but,  unlike the preceding sections, also  depend on the elesticity tensor and on the valence deformation potentials.  They are given  by  
 
\begin{align}\label{wavefunction110}
\setlength\arraycolsep{0pt}
\begin{pmatrix}
|\psi_{1,1}>\\
|\psi_{1,2}> \\
| \psi_{2,1}>\\
 |\psi_{2,2}>\\  
\end{pmatrix} 
 = 
 & \begin{pmatrix}
 -i \overline{M}_1 & 0  &   \overline{L}_1 &0\\
0&  \overline{L}_1 &0&i \overline{M}_1  \\
i \overline{M}_2 & 0  & -   \overline{L}_2&0\\
0  &   -   \overline{L}_2& 0&- i \overline{M}_2\\
\end{pmatrix}
\nonumber \\ &\times
\begin{pmatrix}
|+3/2>\\
 |+1/2>\\
|-1/2>\\
 |-3/2>\\ 
\end{pmatrix}
\end{align}

\noindent
In order to explain the parameters entering in the above equation, we first define $L_0 = \eta   B_{\epsilon} (S_{11}- S_{12})$, $M_1=  \eta D_{\epsilon} S_{44}/2$,   $L_1 = [L_0 ^2+M_1^2]^{1/2} - \eta L_0$. One also has $M_2 (\eta) = M_1 (-\eta) $ , $L_2 (\eta) = L_1 (-\eta) $ and finally  $\bar{M}_j =  M_j / [M_j^2  + L_j^2 ]^{1/2}$ and  $\bar{L}_j =  L_j / [M_j^2  + L_j^2 ]^{1/2}$, where $j=(1,2)$.    With the above expressions,   one finds the matrix elements of   $\hat{p}_{x'}$,$ \hat{p}_{y'}$, and $ \hat{p}_{z'}$,  given in the supplementary material. \\

  \begin{figure}[tbp]
 \includegraphics[clip,width=8 cm] {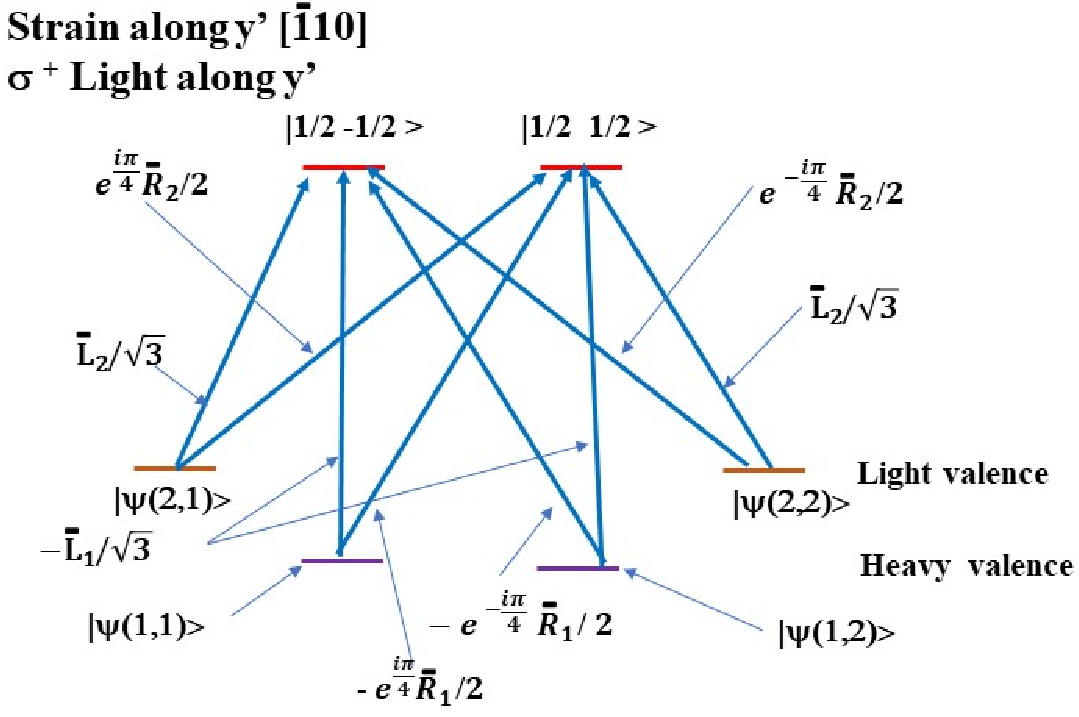}
\caption{Optical matrix elements, in units of  $ \Pi_{cv}$, of the transitions from the $|\psi (i, j)>$ valence wave functions,  for a  strain  and light excitation along the  $[\bar{1}10]$ ($\boldsymbol{y'}$) direction. Here,  the ordering of heavy and light valence levels   corresponds to a compressive strain. The expressions for the quantities $\overline{R}_j$ and  $\overline{L}_j $ are given in the text. }
\label{Fig6}
\end{figure}
\begin{figure}[tbp]
 \includegraphics[clip,width=8 cm] {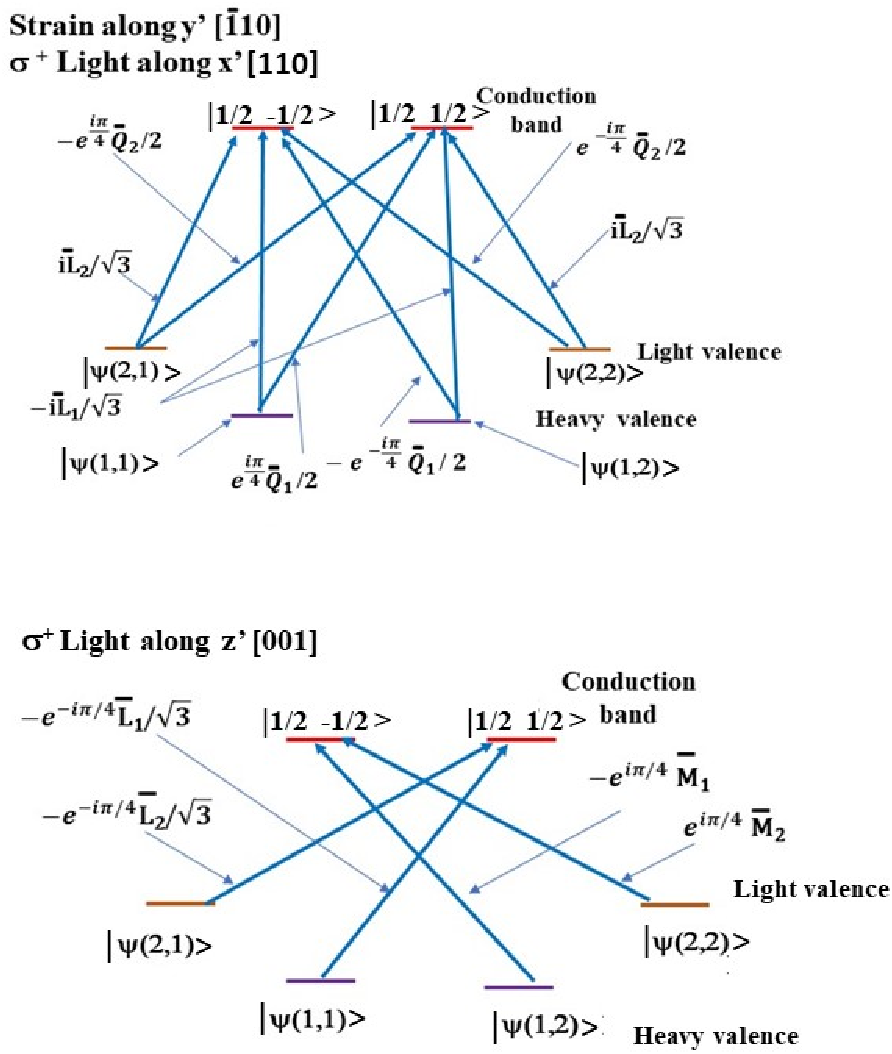}
\caption{Same as Fig. \ref{Fig6}, but for a light excitation along the   $x'$ direction (top panel) and the $z'$ direction (bottom panel), both  perpendicular to the strain direction.  }
\label{Fig7}
\end{figure}

For a $\sigma ^{+}$-polarized  light excitation along the strain direction $y'$, the matrix elements of  $ \hat{p}^{y'}_{+}$ are  given in  Fig. \ref{Fig6}, where $ \overline{R}_j=   \overline{L}_j/\sqrt{3} - \overline{M}_j$.  Using the model of Sec. II, one finds that the  mean electronic  spin generated from  valence level $j$   is given by
 
\begin{equation}\label{spin110y'}
  <S_j>=  \frac{\hbar}{2} P_{j \to c }^{ y'}\boldsymbol{y'}
\end{equation}  
 
\begin{equation}\label{polar110y'}
 P_{j \to c }^{ y'}= -\frac{1- \sqrt{3}\frac{\overline{M}_j }{\overline{L}_j }}{1 + \frac{1}{4} [1-\sqrt{3}\frac{\overline{M}_j}{\overline{L}_j}]^2}
\end{equation}
so that, with the present parameter values and as shown in Fig. \ref{Fig2}, one has $P_{h \,\to\, c}^{ y'}\approx -1$ and   $P_{l \,\to\, c}^{ y'} \approx 1$.  These results  depend mostly on the shear tensor components since, for $S_{44}=0$, one has $M_j=0$ and therefore   $ P_{j \to c }^{ y'}=1$. \ 
 
The quantity   $P_{j \to c }^{ y'}$ is also  sensitive to changes of the valence deformation potentials. This is seen in Fig. \ref{Fig2} where  the values between parentheses were obtained using slightly  modified values  of $B_{\epsilon } =-1.7$ $eV$ and  $D_{\epsilon } =-5.4$ $eV$ that is, within the uncertainty range given by  \cite{ivchenko1997}.  The   overall spin polarization, including all transitions but neglecting the effective mass correction is obtained using  Eq. \ref{polartot 110y'}.  After some simple algebra, one finds  $ P^{[\bar{1} 10]}_{y'}=-1/2$, as in the previous cases where light propagates along the strain direction. Remarkably, this exact result depends neither on the stress tensor coefficients nor on the valence deformation potentials. \

\noindent

 For light along the $z'$ direction,   i. e. perpendicular to the strain, the matrix elements of  $\hat{p}_{+}^{z'}$ are given   in the bottom panel of Fig. \ref{Fig7}. Since the density matrix is diagonal, the spin polarization of conduction electrons generated from valence level $j$  lies along the direction of light excitation and is of the form \ 

\begin{equation}\label{spin110z'}
  <S_j>=  \frac{\hbar}{2}P_{j \to c }^{ z'}\boldsymbol{z'}
\end{equation}  

\noindent
where  the spin polarization of the conduction electrons is given by 
\begin{equation}\label{polar110x'}
P_{j \,\to\, c }^{ z'}= \frac{1- \frac{3\overline{M}_j^2 }{\overline{L}_j^2 }}{1 +\frac{3\overline{M}_j^2}{\overline{L}_j^2}}
\end{equation}

 The obtained polarization differs significantly from the case  where light propagates parallel  to the strain direction $y'$ since, as shown in  Fig. \ref{Fig2}, the polarization from the heavy valence band becomes slightly positive for a $\sigma^+$ excitation. The exact value depends on the experimental parameters used for valence  deformation potentials and stress tensor. On the other hand, the polarization generated by the light valence band  is almost identical to the corresponding value for the   $[0 0 1]$ strain direction and light excitation perpendicular to the strain.  In the same way as for light excitation along the strain direction, the polarization value depends on the shear deformation since, for $S_{44}=0$, one has  $ P_{j \,\to\, c y'}=1$. Finally, the overall electron spin polarization, including all transitions, is equal to $ P^{[\bar{1} 10]}_{y'}= -1/2$, an exact result independent on the stress tensor and on valence deformation potentials. \

Light excitation perpendicular to the strain direction can also occur along the  $x'$  directions. The polarization differs as well from the one of the other orthogonal directions $ z'$ and  $x'$. This is expected, since the representation corresponding to the $ z'$,$ x'$,$y' $ directions all differ in the $C_{2v}$ symmetry group.   For light along the $x'$ direction,  the matrix elements of  $\hat{p}_{+}^{x'}$ are given   in the top  panel of Fig. \ref{Fig7}. The spin polarization of conduction electrons generated from valence level $j$  is of the form \ 

\begin{equation}\label{spin110z'}
  <S_j>=   \frac{\hbar}{2} P_{j \to c}^{x'}\boldsymbol{x'}
\end{equation}  

\noindent
where  the spin polarization of the conduction electrons is now given by 
\begin{equation}\label{polar110z'}
 P_{j \,\to\, c}^{x'}= -\frac{1+ \sqrt{3}\frac{\overline{M}_j }{\overline{L}_j }}{1 + \frac{1}{4} [1+\sqrt{3}\frac{\overline{M}_j}{\overline{L}_j}]^2}
\end{equation}

As seen from comparison of the top and bottom panels  Fig. \ref{Fig7},  the matrix elements of  $\hat{p}_{+}^{x'}$  are quite distinct for light along the $z'$ and $x'$ directions. Fig. \ref{Fig2} shows that  this is also the case for the polarization values. Note that, for  excitation along the   $x'$ direction,  as well as for the other cases considered in Fig. \ref{Fig2}, the overall conduction spin polarization is $ P_i= - 1/2$, a remarkable analytical result.   \
 
 \section{Interpretation : Universality of  $P_i$}
Fig. \ref{Fig2} shows that, in spite of the strong dependence of the oscillator strengths and of the spin polarizations $P_i^h$ and  $P_i^l$ on experimental conditions,  the overall conduction electron spin polarization  $P_i$  is equal   to  $ -0.5$ in all cases,  independently of the deformation potentials and shear tensor components.  This universal behavior can be interpreted  by considering  a gedanken experiment in which, along with the transitions from the heavy and light valence levels,  one also excites the  optical transitions from the spin-orbit split-off band, leading to an initial spin polarization  $P_i ^{so}$,  an initial electron concentration proportional to $r_{so}$ and a total spin polarization including all transitions $P_i^{tot}$. Extending  Eq. \ref{polartot 110y'}, one finds 
   \begin{equation}\label{poltot}
[ r_{h}+  r_{l}+ r_{so}] P_i^{tot} =  [r_{h}+  r_{l}]  P_i  + r_{so}   P_i ^{so}  
\end{equation}
  
Since the conduction electron spin polarization generated by all the transitions from the valence band, including the spin-orbit split-off band,  does not depend on the spin -orbit splitting and should be zero,  one has $   P_i  = -  P_i ^{so} r_{so}/ [r_{h}+  r_{l}]$. Using the facts that the wavefunctions of the spin-orbit split-off band transform like the $\Gamma_7$ representation and that the strain hamiltonian has the spherical symmetry both for $\Gamma_6$ conduction electrons and $\Gamma_7$ spin-orbit valence states, one finds that  $P_i ^{so}=1$ for any strain/light- propagation directions configuration. This result is in agreement with the atomic and spherical models, as can be found from a calculation of the trace of the generation amtrix  $ r_{so} =2\Pi _{cv}/3$  \cite{note30}.  Since it can also  be shown that $r_{h}+  r_{l}=4\Pi _{cv}/3$ independently on the direction of strain and light excitation, one obtains  indeed $   P_i  =-0.5$. This proves the universal nature of $   P_i $\

 \section{Conclusion}
  
In the present work,  we obtain the values of the optical matrix elements and of the initial polarizations for transitions from heavy and light  valence levels using a simple analytical model.  We  consider the case of a strain along the $[0 0 1]$, $[1 1 1]$ or $[\bar 1 1 0]$ directions and a circularly-polarized light excitation along the direction of the strain and along the perpendicular direction. We include the strain-induced modification of the valence wavefunctions, which was overlooked within both the atomic model and the spherical approximation. \

 It is found that,  under compressive strain, the light valence level lies above the heavy valence one.  The wavefunctions, as well as the optical oscillator strengths, do not depend on the value of the uniaxial strain but only on its sign since, upon changing of this sign,  each state (heavy or light) has  unchanged matrix elements. While  the conduction electron mean spin  may, quite generally,  lie out of the  direction of light excitation, this is not the case for the above particular strain directions, for which this mean spin lies along the light excitation direction.  As shown  in Fig. \ref{Fig2} for a compressive strain, three different situations can be outlined 
 
 a) Light excitation parallel to the direction of the uniaxial strain  :  For electrons generated from both heavy and light valence bands the initial polarizations are identical for  the  $[0 0 1]$, for  the $[1 1 1]$  strain and for the   $[\bar 1 1 0]$ directions and coincide with the predictions of the atomic model. This  implies that, in this case, the  strain-induced modifications of the valence wavefunctions do not strongly affect the spin polarization. 

b) Light excitation perpendicular  to a strain along $[0 0 1]$ and  $[1 1 1]$ : Here, the atomic approximation fails.  As seen from    Eq. \ref{depstrain},  the polarization values   coincide with the predictions of the spherical approximation, including the zero value of the polarization generated from the heavy valence levels.  This implies that, as for the preceding case, the strain-induced modifications of the wavefunctions have no effect on the conduction electron spin orientation. \

c)  Light excitation perpendicular  to a strain along $[\bar 1 1 0]$ : Here  the atomic approximation and the spherical approximation both fail. Remarkably i) the matrix elements and polarization values are not the same for light along the two directions perpendicular to the strain, the $x'$ and $z'$ directions, both for transitions from the heavy- or from  the light-valence band. ii) The polarization values for transitions from heavy or light valence levels depend on   both the elasticity and valence deformation potential values.. The spin polarization of electrons generated from the heavy valence level can be zero or slightly positive depending on the exact values of these tensors.  iii) The polarization for light propagating perpendicularly to the quantization axis z', i.e. along the x' (orthogonal to the strain) or y' (parallel to the strain) differs, as expected from group theory.\

  It is remarkable to notice that, in the case of a strain along   $[\bar 1 1 0]$   and for a light excitation along the $x'$  and  $z'$ directions,   the overall  electron spin polarization is exactly equal to $-1/2$ and is independent on the elasticity tensor.  As shown  in Fig. \ref{Fig2} and  interpreted in the preceding section, this result is general and valid for all directions of strain and light excitation. \ \

\appendix

\section{Optical pumping and time reversal}
\label{appendix}

The change of the excitation light helicity induces the change of the linear momentum operator $\hat{p}_+$ by  $\hat{p}_-$ , which is linked to $\hat{p}_+$  by time reversal, according to     $\hat{p}_{\mp}^{\zeta} =  \hat{K}^{\dagger} \hat{p}_{\pm}^{\zeta}  \hat{K} $, as can be found using 
 $ \hat{K}^{\dagger}  \hat{ p}_{\alpha}  \hat{K}= - \hat{p}_{\alpha}$ and the definition  \cite{koster1963} of  $\hat{p}_{\pm} $.  Defining  $\sigma= 3/2 -\tau$, as the quantum number of the heavy and light hole pseudospins and using the antilinear properties of the time reversal operator  \cite{littlejohn2021,   messiah1962}, one can write 
  
 \begin{equation}\label{elmat}
\mathscr{A}_{\pm}^{\zeta} (s; j, \sigma)= (-1) ^{1-\sigma-s}{ <\psi_{c,-s}| (\hat{K}^{\dagger} \hat{p}_{\pm}^{\zeta} \hat{K} ) | \psi _{j, - \sigma}>}^*
\end{equation}
 where $j= 1,2$ and $\sigma = \pm 1/2$. One readily obtains after some manipulation  
 \begin{equation}\label{elmat2}
\mathscr{A}_{\pm}^{\zeta} (s, j, \sigma)= (-1) ^{\sigma-s}\mathscr{A}_{\mp}^{\zeta} (-s, j, -\sigma)^*
\end{equation}

 One has then 
\begin{equation}\label{matel2}
 \mathscr{A}_{\pm}^{\zeta} (s, j, \tau)= (-1)^{3/2-\tau-s}\mathscr{A}^{\zeta *}_{\mp} (-s, j, \overline{\tau}) 
\end{equation}

\noindent
where $ \tau,  \overline{\tau} =1,2$ and  $ \tau \neq \overline{\tau}$.  As a result,  a change of light helicity induces a change of sign of the conduction electron spin polarization.\

\acknowledgments 
{  }
 
\bibliographystyle{apsrev}
%\begin{thebibliography}

%\bibliography{cadizref}

%\bibliography{nanofil theor v29.bbl} 

\begin{thebibliography}{30}
\bibitem[{not({\natexlab{c}})}]{note29}
	\bibinfo{note}{The corresponding values for electronic levels in the valence
		band are equal to $ A_{\epsilon} = 8 eV$ , $ B_{\epsilon}= 2 eV$ and $
		D_{\epsilon}=5 eV$ respectively.}
\bibitem[{not({\natexlab{e}})}]{note27}
	\bibinfo{note}{It is assumed that the interband spin relaxation time is shorter
		than both the intraband relaxation time and the hole population relaxation
		time between the upper and lower hole bands. As a result, no interband spin
		coherence occurs in the model.}
\bibitem[{not({\natexlab{f}})}]{note20}
	\bibinfo{note}{We neglect here matrix elements of the type
		$<\mathscr{X}|\hat{p}_y |\mathscr{Z}>$ which are small with respect to
		$\Pi_{cv}$ in weakly polar cubic semiconductors such as GaAs}.
\bibitem[{not({\natexlab{g}})}]{note22}
	\bibinfo{note}{The situation is the same for a tensile strain ($\eta=1$) since
		in this case, the the matrix elements of $p_z$ for transitions from the heavy
		valence levels $\psi _{2,1}$ and $\psi _{2, 2}$, lying above the light
		valence levels, are zero}.
\bibitem[{not({\natexlab{h}})}]{note23}
	\bibinfo{note}{This result originates more generally from the symmetry of
		wavefunctions and operators in the $C_{3v}$ point group. Using the
		compatibility table between $T_d$ and $C_{3v}$, we see that, in $k=0$, the
		heavy valence states belong to two one-dimensional representations $\Gamma_5$
		and $\Gamma_6$, mutually conjugated. The conduction states belong to the
		two-dimensional representation $\Gamma_6$. Finally $p_Z$ and $A_Z$ belong to
		the $\Gamma _1$ scalar representation. It can be seen from the coupling table
		that $< \pm 1/2| p_Z| \psi _{1, j} >=0$.}
\bibitem[{not({\natexlab{i}})}]{note30}
	\bibinfo{note}{The total hamiltonain, including kinetic and strain induced
		terms has a spherical symmetry both for $\Gamma_6 $ conduction electrons and
		$\Gamma_7$ spin-orbit valence states, since it depends only on $ k^2$ and
		$\epsilon$ (i.e. the hydrostatic term). Hence the spin-orbit eigen states
		defiined for $k=0$ and unstrained materials are not modified by strain.}
\end{thebibliography}
\end{document}